\documentclass[prb,aps,superscriptaddress,showpacs]{revtex4}
\usepackage[dvips]{graphicx}
\usepackage{amsmath}
\usepackage{epsfig}

\def\P{{\bf{P}}}
\def\M{{\bf{M}}}
\def\E{{\bf{E}}}

\def\B{{\bf{B}}}

\begin{document}

\title{ Multiferroic Thermodynamics  }

\author{G.R.Boyd}
\author{P.Kumar}
\affiliation{Department of Physics, University of Florida, PO Box 118440
    Gainesville, FL 32611, USA}
\author{S.R.Phillpot}
\affiliation{Department of Materials Science and Engineering, University of Florida,
    Gainesville, FL 32611, USA}

\begin{abstract}
We have studied the thermodynamic properties of a multiferroic that couples ferromagnetic and ferroelectric order. Some of the results are independent of the form of the free energy. We calculate the temperature dependence of the electric, magnetic, and magnetoelectric susceptibilities. The cross susceptibility has a temperature dependence related to the mixed (with respect to E and B) derivatives of the specific heat. The phase transitions are all second order. In particular, the phase boundary T$_M$(E), where T$_M$ is the lower magnetic transition as a function of electric field, is described by the Ehrenfest relation. The magnetoelectric susceptibility is nonzero only below the lower of the two transition temperatures. We study the properties of the specific heat, with and without the inclusion of gaussian fluctuations. The perturbative renormalization group is used to understand the fixed points of the theory, and we include a discussion of the effect inhomogeneities have for this model.
\end{abstract}
\date{\today}

\maketitle

\section{Introduction}

Multiferroics couple polarization, magnetization, and the elastic response in a material. The subject has been considered interesting for some time now. Interest in the electric control of the magnetic properties of a system for applications requires the magnetic and polar orders be coupled as strongly as possible. There are materials that exhibit multiferroic properties and have been subject of several reviews recently\cite{AdvPhysReview,Mostovoy,Ramesh}. Many physical systems can be described by the interplay of more than one order parameter, but it is the new context of multiferroics which leads back to this classic problem.

Dzyaloshinskii\cite{Dzyaloshinskii} first pointed out that, on symmetry grounds, the coexistence antiferromagnetic and ferroelectric order in Cr$_2$O$_3$ was possible.  Astrov\cite{Astrov} confirmed the prediction for Cr$_2$O$_3$ by measuring the magnetization due to an electric field. Rado et al\cite{FolenRado} specifically demonstrated the nature of the interaction between the magnetic and electric degrees of freedom by reporting that the temperature dependence of the cross susceptibility (defined below) was different depending on the relative orientation of the magnetic field and the staggered magnetization in the antiferromagnetic Cr$_2$O$_3$. During the 1960s and 1970s several multiferroic materials were discovered, such as BiFeO$_3$, BiMnO$_3$, and the boracite Ni$_{3}$B$_{2}$O$_{13}$I. A review by Smolenskii and Chupis\cite{SmolenskiiChupis} is a compilation of the subject until the early 1980s. Recently, the discovery of a new class of multiferroics, notably TbMnO$_3$ \cite{kimura_goto_shintani_ishizaka_arima_tokura_2003}, reignited interest in multiferroic research. A broad class of these multiferroics follows the chemical formulas, RMnO$_3$ and RMn$_2$O$_5$\cite{hur_park_sharma_ahn_guha_cheong_2004} , where R is a rare earth ion. The appearance of spiral magnetic order causes a polarization. In this more recent class of multiferroics, the ferroelectricity is referred to as improper meaning that the primary order, the magnetism, is said to induce the electric polarization which occurs as a consequence of the symmetry. Mostovoy\cite{Mostovoy} provided a general argument based on symmetry and Ginzburg-Landau theory for why this happens. This is to be contrasted with proper ferroelectrics, where a spontaneous polarization exists of its own accord. We may then classify the macroscopic view of this behavior as follows: in an improper ferroelectric T$_M$ and T$_E$ coincide ($\P$ occurs as soon as the appropriate $\M$ does) while in a proper ferroelectric T$_E$ is independent from (usually greater than) T$_M$. This paper is a study of the later case, its properties and fixed points of the theory.

The physics of two coupled order parameters has been studied in various contexts, see \onlinecite{Imry,Watanabe,Aharony,Kosterlitz} and references therein. More recently, the dynamics and domains in ferroelectric ferromagnets have also been studied\cite{Houchmandzadeh,Sukhov,Scott}. We consider the thermodynamic response of a system which does not depend on any specific form of the free energy, then specialize to ferroelectric-ferromagnetic order to investigate the temperature dependence of the susceptibility which, to our knowledge, does not appear in previous work. An outline of our paper is as follows: we begin with a catalog of thermodynamic results including Maxwell relations and a bound on the magnetoelectric effect\cite{BrownMEbound}. In the next section, we specify the simplest form for the free energy and examine the susceptibilities at both transition temperatures. We find that in the absence of a fluctuation-induced response, the magnetoelectric susceptibility is only nonzero when both $\P$ and $\M$ are nonzero. Finally we examine the fixed points for the theory and the effect of inhomogeneities.

\section{Thermodynamics}
\subsection{Maxwell Relations}

The Helmholtz $F(\P, V, \M, T)$ and the Gibbs $G(\E, \P, \B, T)$ free energies are functions of the
volume V, magnetization $\M$, temperature T and Electric field $\E$, polarization $\P$, pressure P, and magnetic field $\B$. They are described by,

\begin{equation}
dF=-SdT-PdV+ \E d\P + \B d\M
\end{equation}
\begin{equation}
dG=-SdT+V dP- \P d\E - \M d\B
\end{equation}

These are the definitions of thermodynamic observables as derivatives of the free energy.
By noting the equality of cross derivatives, it follows that,

\begin{equation}
\frac{\partial \M}{\partial \E}|_B=\frac{\partial \P}{\partial \B}|_E
\end{equation}

Which is the reciprocity statement for the cross susceptibility $\chi_{12}$  in Eq. \ref{susc}.
By looking at the other cross derivatives, the following relations can be derived:

\begin{equation}
\frac{\partial^2 C}{\partial \E^2} = T \frac{\partial^2 \chi_E  }{\partial T^2} \; \; \frac{\partial^2 C}{\partial \B^2} = T \frac{\partial^2 \chi_{M}}{\partial T^2}
\end{equation}and
\begin{equation}
\frac{\partial C}{\partial \B \partial \E} = T \frac{\partial^2 \chi_{12}}{\partial T^2}
\end{equation}
\begin{equation}
\frac{\partial^2 \chi_{12}}{\partial  P^2}= \frac{\partial^2 (V \kappa)}{\partial \B \partial \E},  \;\;
\frac{\partial^2 \chi_{E}}{\partial  P^2}= \frac{\partial^2 (V \kappa)}{ \partial \E^2}, \;\;
\frac{\partial^2 \chi_{M}}{\partial  P^2}= \frac{\partial^2 (V \kappa)}{ \partial \B^2}
\end{equation}

Here $\chi$ is the differential susceptibility (the subscript E for electrical  and  M for magnetic),
$\kappa$ is the compressibility, V is the volume and C the specific heat at constant volume.

The mechanical stability of the ground state also leads to the inequality
\begin{equation}
\frac{\partial^2 F}{\partial \M^2}\frac{\partial^2 F}{\partial \P^2}-(\frac{\partial^2 F}{\partial \M \partial \P})^2 \leq 0
\end{equation}
which in turn implies that
\begin{equation}
\chi_{12}^2 \leq \chi_e \chi_m
\end{equation}
This inequality was first derived by Brown et al\cite{BrownMEbound} and remains an essential testing ground against unsupportable approximations.

The Ehrenfest definition of the order of a phase transition needs to be considered carefully where there is more than one mechanical field. The thermodynamic phase boundaries for a first order transition and a single order parameter are described by the Clausius-Clapeyron equation. For a second order phase transition, the corresponding equation is the Ehrenfest equation\cite{Ehrenfest}.

The Clausius-Clapeyron equations for T$_M$(E), which can be obtained from $dG_i=-S_i dT +V dP - P dE - M dB$ by choosing two points on either side of the critical line in the T-B plane, takes the form:
\begin{equation}
\frac{\partial T_M}{\partial \E} =-\frac{\P_1-\P_2}{S_1-S_2}
\end{equation}
For completeness, we present all four Ehrenfest equations with the two mechanical fields, $\E$ and $\B$:
\begin{equation}
\left(\frac{\partial T_E}{\partial \E}\right)^2 =\frac{T_E \Delta \chi_E}{\Delta C}, \;\;\; \left(\frac{\partial T_M}{\partial \E} \right)^2=\frac{T_M \Delta \chi_E}{\Delta C}
\end{equation}
\begin{equation}
\left(\frac{\partial T_E}{\partial \B} \right)^2=\frac{T_M \Delta \chi_M}{\Delta C}, \;\;\; \left(\frac{\partial T_M}{\partial \B} \right)^2=\frac{T_M \Delta \chi_M}{\Delta C}
\end{equation}

In a ferromagnet (or ferroelectric) in any finite field there is no phase transition. And yet, if the transition is second order at a finite field, as it can be for an antiferromagnet, the evolution of T$_c$ should be described by the Ehrenfest equations above. When the two order parameters are coupled, as we will show below, $\Delta \chi_M$ is zero at $T_E$ but $\Delta \chi_E$ is non-zero at $T_M$. The ferromagnetic transition at $T_M(\B,\E)$ is second order in $\E$, unlike the uncoupled result.

\subsection{Adiabatic Processes}

Adiabatic processes are used in cooling, in a Joule-Thompson decompression or demagnetization at low temperatures.
In a multiferroic, the entropy S(\E, \B, P, T) depends on the external fields.  The cooling arising from an adiabatic process is described by dS=0 which upon substituting the Maxwell relations becomes:
\begin{equation}
\frac{d\P}{dT} d\E+  \frac{d \M}{dT}d\B +\frac{dV}{dT} dP+\frac{C_v}{T}dT=0
\end{equation}
\begin{equation}
\frac{dT}{dE} =  -\frac{T}{C_v} \frac{d\P}{dT}
\end{equation}
We will defer any explicit calculation of the integral for the which depends on the explicit temperature dependencies of the specific heat and also of the order parameter.

\section{Free Energy Functional}

The free energy of a system with more than one vector order parameter can be written as:
$F=F_E+F_M+F_{int}$
\begin{equation}
\label{Fe}
F_E= \frac{\P^2}{2 \chi_{E0}} + b_E \P^4
\end{equation}
\begin{equation}
\label{Fm}
F_M= \frac{\M^2}{2 \chi_{M0}} + b_M \M^4
\end{equation}
\begin{equation}
\label{FI}
F_{i}= k(\M^2 \P^2)
\end{equation}
\begin{equation}
\chi_{E0}^{-1}=a_{E0}\left( \frac{T}{T_{E0}}-1\right)
\end{equation}
\begin{equation}
\chi_{M0}^{-1}=a_{M0} \left( \frac{T}{T_{M0}}-1 \right)
\end{equation}

We are considering here a ferromagnet described by $F_M$ and a ferroelectric represented by $F_E$. The interaction between the two degrees of freedom is given by $F_i$. The possible inhomogeneous instabilities will be considered in the summary section below. The coefficients of the free energy here are all constants with the exception of the various susceptibilities which are assumed to be described by the usual Curie-Weiss law for localized moments. For specificity we take $T_{E0} >T_{M}$ as the transition temperatures respectively for the electric polarization and the magnetization.  In general, the interaction between the magnetization and electric polarization must be a scalar and could be in the form $k_1 \M^2\P^2 + k_2 (\M.\P)^2$ which is required by the time reversal invariance. This term determines the relative angle between $\M$ and $\P$. Once the relative angle is determined, the free energy becomes of the form in Eq.\ref{FI} with its coefficients renormalized. There is another motivation for considering a biquadratic interaction of this form, beyond it being the simplest consistent with symmetry. If we think of the exchange for local moments and expand,  since the dipole moment is classically proportional to the displacement, the coupled elastic and Heisenberg interaction, through $\nabla J(x^0_{ij}) P M^2 $, result in a $\P^2 \M^2$ form. While this argument is only heuristic, it informs the choice of free energy.

The equations of state are the conventional thermodynamic equations for a Helmholtz energy:
\begin{equation}
\E=\frac{\partial F}{\partial P}
\end{equation}
\begin{equation}
\B=\frac{\partial F}{\partial \M}
\end{equation}
For a homogeneous system, the free energy above is quite sufficient to derive all thermodynamic properties.
 In a multiferroic, the linear response includes cross susceptibilities $\chi_{12}$, defined by

\begin{equation}
\label{susc}
\begin{array}{cc}
\delta \M =& \chi_M \B + \chi_{12} \E \\
\delta \P =& \chi_{12} \B + \chi_E \E \\
\end{array}\end{equation}

The free energy functional in the conventional form includes the external fields $-\P.\E$ and $-\M.\B$,
and thus the ground state is a minimum of f. The dimensionless free energy $f=F/F_E(0)$, is written in terms of: $p=P/P_0$, $m=M/M_0$, $\ell=F_M(0)/F_E(0)$, $\epsilon=\frac{ E}{F_E(0)} P_0$, $\beta.m \ell= \B.\M /F_{E}(0)$.

\begin{equation}
f = \frac{F}{F_{E0}}= f_e+f_m+2k m^2 p^2=
2 (\frac{T}{T_{E0}}-1)p^2 +  p^4 -p. \epsilon
+ \ell(2(\frac{T}{T_{M0}}-1)m^2 + m^4 - m.\beta)
+  2 k (m^2 p^2)
\label{freeenergy}
\end{equation}

The equations of state in the scaled variables becomes,

\begin{equation}
\frac{\epsilon}{4}= (\frac{T}{T_{E0}}-1) p +  p^3 +  k m^2 p
\end{equation}

\begin{equation}
 \frac{\beta}{4}=(\frac{T}{T_{M0}}-1) m +  m^3 +  \frac{k}{\ell} m p^2
\end{equation}

\subsection{Order Parameter}

We consider $T_{E0}>T_{M0}$. The solutions are:

\begin{equation}
\begin{array}{ccc}
      p_0=m_0=0 & & T>T_{E0}>T_M \\
      m_0^2=0, \; p_0^2(T) = (1-\frac{T}{T_{E0}}) & & T_{E0}>T>T_M. \\
      m_0^2(T) = \frac{1-\frac{k}{\ell}}{1-\frac{k^2}{\ell}}(1-\frac{T}{T_M}),\; p_0^2(T) = \frac{1-k}{1-\frac{k^2}{\ell}}(1-\frac{T}{T_E})  & & T_{E0}>T_M>T\\
      \frac{T_E}{T_{E0}}=\frac{1-k}{1-k\frac{T_{E0}}{T_{M0}} } & &  \frac{T_M}{T_{M0}}=\frac{1-k/\ell}{1-\frac{k}{\ell}\frac{T_{M0}}{T_{E0}} }
\end{array}\end{equation}

\begin{figure}[ht]
\includegraphics[width=90mm]{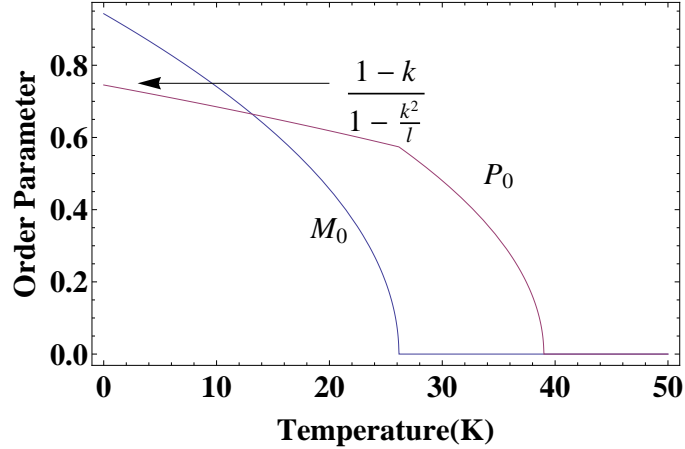}
\caption{(Color Online) The polarization and magnetic moment versus temperature. P shows a jump at $T_M$ when M condenses. }
\label{PMvT}
\end{figure}

As shown in Fig. \ref{PMvT}, the mean field order parameter $\P$ appears at $T_{E0}$ continuously.

Since we have chosen $T_{E0}/T_{M0} >1$, we have $T_E/T_{E0} >1$ but $T_M/T_{M0} < 1$.  The electrical transition takes place at T$_{E0}$, which is unchanged by the interaction, and T$_{M0}$ is re-normalized by interactions. The physical magnetic transition takes place at T$_M$, not T$_{M0}$.  As shown in Fig. \ref{PMvT}, there is a kink in p$_0$ at T$_M$. This kink is due to the fact that the scale temperature for  $p_0$, for $T< T_M$ is $T_E >T_{E0}$.

\subsection{Susceptibility}

\begin{figure}[htbp]
\begin{center}$
\begin{array}{cc}
\includegraphics[width=90mm]{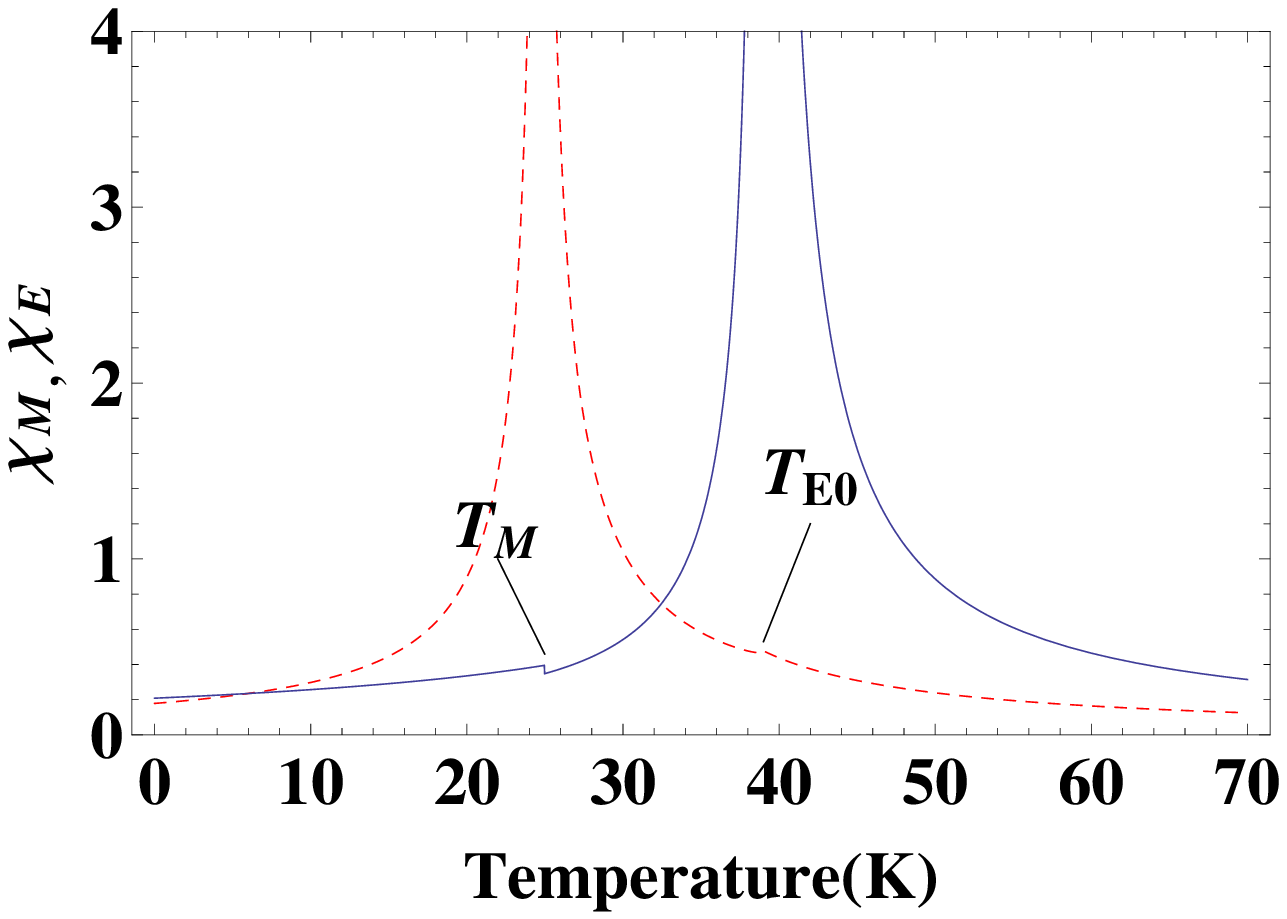} &
\includegraphics[width=90mm]{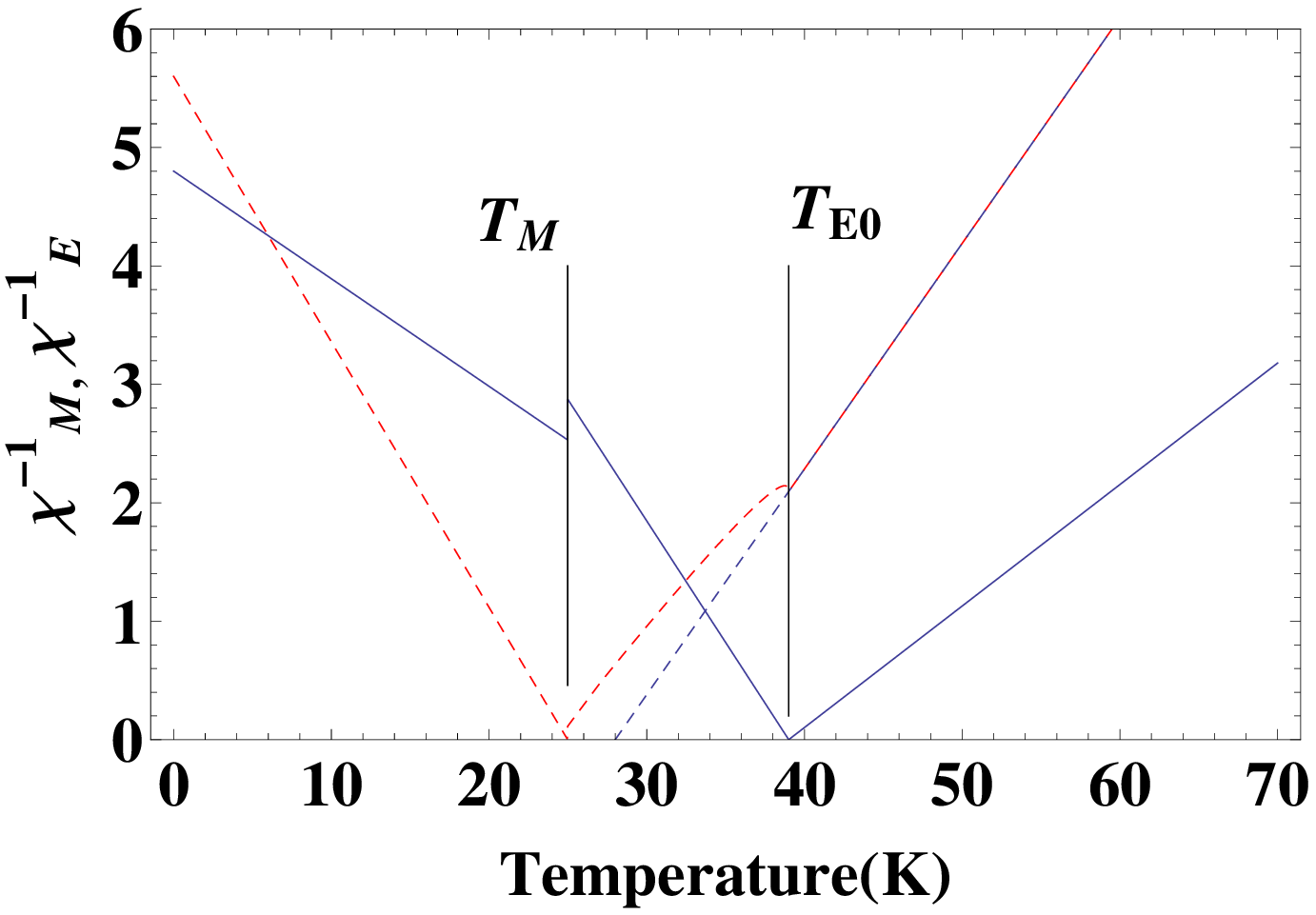}
\end{array}$
\end{center}
\caption{(Color Online)  (Left) The ferroelectric susceptibility, the blue solid line, and the magnetic susceptibility, the dashed red line. (Right) The inverse susceptibilities, the blue solid line, $\chi_E^{-1}$, and the thin dashed line representing the inverse magnetic susceptibility, $\chi_M^{-1}$. An auxiliary dashed line is also present to underscore the slope change in $\chi_M^{-1}$ at T$_{E0}$.
}
\label{MAG}
\end{figure}

\begin{figure}[ht]
\includegraphics[width=90mm]{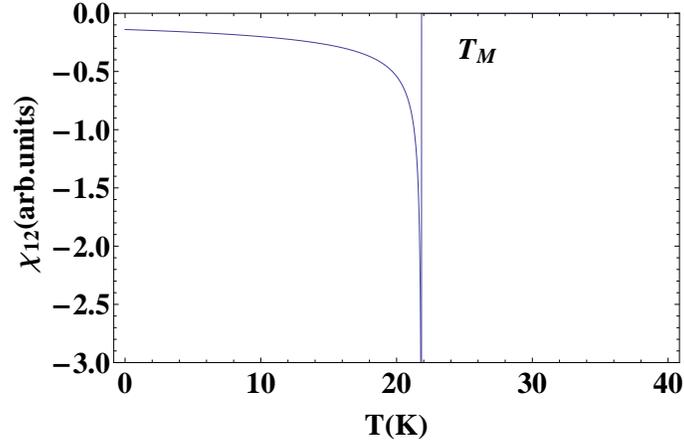} 
\caption{(Color Online) The cross-susceptibility becomes nonzero only below $T_M$ and has the opposite sign of the other susceptibilities.  }
\label{Chi12}
\end{figure}

Following the equation of state, we can derive the electric, magnetic, and the cross susceptibilities as defined in Eq. \ref{susc}.
They both take a Curie-Weiss form, also have a structure at the "other" transition temperature.  Figures (\ref{MAG}-\ref{Chi12}) show these features for a specific choice of the parameters(k$=.4$,$\frac{k}{\ell}=.3$). The cross susceptibility diverges at $T_M$. As expected, it vanishes above $T_M$. This excludes the possibility of a fluctuation induced response, which we neglect in our study. The ferroelectric susceptibility, exhibits a divergence at its transition temperature and an anomaly at the magnetic transition temperature. The magnetic susceptibility demonstrates paramagnetic behavior above it's transition and diverges at $T_M$. It shows a cusp at the ferroelectric transition temperature coinciding with the smooth onset of $\P$. The inverse electric susceptibility, $\chi_E^{-1}$, is zero at its transition temperature and jumps at the magnetic transition temperature. The size of this jump depends on the energetics of the ferroelectric and magnetic free energies, and occurs because both D and the numerator vanish as  T approaches T$_M$ from below. The inverse magnetic susceptibility, $\chi_M^{-1}$, is zero at $T_M$, and shows a change at the ferroelectric transition temperature due to the onset of $\P$, resulting in a cusp.

The formal expressions are,

\begin{equation}
\chi_E(T>T_{E0}) = \frac{1}{4(\frac{T}{T_{E0}}-1)}, \;\;\;\; \chi_E(T_M<T<T_{E0}) = \frac{1}{8( 1-\frac{T}{T_{E0}} ) }
\end{equation}

\begin{equation}
\chi_E(T<T_M) = \frac{1}{4D}( ( \frac{T}{T_{M0}}-1)+3 m_0^2 +  \frac{k}{\ell} p_0^2 )
\end{equation}

\begin{equation}
\chi_M(T>T_M) = \frac{1}{4 \ell (( \frac{T}{T_{M0}}-1) + \frac{k}{\ell} p_0^2)}
\end{equation}

\begin{equation}
\chi_M(T<T_M) = \frac{1}{4D} ( ( \frac{T}{T_{E0}}-1) + 3  p_0^2 +  \frac{k}{\ell} m_0^2)
\end{equation}

\begin{equation}
\chi_{12} = \frac{1}{4D}(-2 \frac{k}{\ell} m_0 p_0)
\end{equation}

with
\begin{equation}
D=[ (\frac{T}{T_{E0}}-1) + 3 p_0^2 +  k m_0^2][ ( \frac{T}{T_{M0}}-1)+3  m_0^2 +  \frac{k}{\ell} p_0^2 ] -\frac{(2 k m_0 p_0)^2}{\ell}
\end{equation}

 The results have been plotted in Figs. (\ref{MAG}-\ref{Chi12}) for certain specific values of the parameters representative of the general behavior. These parameters include the two bare transition temperatures(T$_{E0}$=39K, T$_{E0}$=28K), the interaction parameter k and the free energy ratio $l$. In all of the results discussed so far, $k^2 < l$.  The ground state is different in the opposite case and leads to a  correspondingly different phase diagram. The cross susceptibility $\chi_{12}$ is non-zero only below T$_M$.

\subsection{Specific Heat}
The free energy describes two second order phase transitions which are coupled.  In a mean field like analysis, that corresponds to two discontinuities at T$_M$ and T$_{E0}$.  The algebraic results are:
\begin{equation}
\frac{C_V}{F_0}= 0 \;\;\;T>T_{E0}, \;\;\; \frac{2T}{T_{E0}^2} \;\;\; T_M<T<T_{E0}, \;\;\; T(\frac{2}{\ell-k^2})(\frac{1}{T^2_{M0}}+\frac{\ell}{T_{E0}^2}-\frac{2k}{T_{M0} T_{E0}}) \;\;\; T<T_M
\end{equation}

\begin{figure}[ht]
\includegraphics[width=90mm]{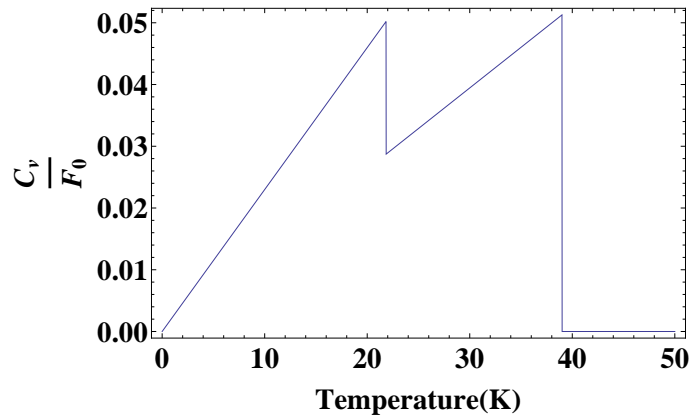}
\caption{(Color Online) The specific heat with the same parameter choice as the susceptibility. }
\label{Cv}
\end{figure}

\section{Inhomogeneous Effects}
There are three gradient terms in the free energy, one for each of the order parameter as well as one in the interaction term.
In the following we first consider the stability of a uniform ground state with respect to inhomogeneous perturbations.
This is followed by a discussion of the fluctuations.

A general interaction between P and M, involving space gradients (and a scalar with respect to inversion and time reversal) can be written as  $P_i M_j M_k \partial_l$.
The gradient can operate on either M or P.
The four indices (i,j,k,l) should be contracted to turn the interaction in to a scalar.  Thus the interaction terms are
\begin{equation}
z_1 M.((M.\nabla)P) + z_2 M^2 (\nabla.P) + z_3 (P.M)(\nabla.M)+z_4 P.[(M.\nabla)M]
\end{equation}
Since all terms are quadratic in M, there is a shift in T$_M$ proportional to the gradient of P. The first pair of terms can be integrated by parts and turned into the second pair. The difference is a total derivative that depends on the boundary properties. The last two terms, involving derivatives of M, are also known as Lifshitz invariants and were introduced by Mostovoy in the present context\cite{Mostovoy}. These terms are linear in P and therefore spontaneously brake the symmetry and make a nonzero P. In other words, as noted by Mostovoy, we have an effective local electric field $E_{int}$ here, proportional to the magnetization gradients.
\begin{equation}
E_{int}=z_3 M(\nabla.M)+z_4 [(M.\nabla)M]
\end{equation}
Which leads to a polarization $P=\chi_E E_{int}$.  The overall effect back on the magnetization (within a local response)
can be summarized as a free energy for magnetization:
\begin{equation}
F=\frac{\M^2}{2\chi_M}-\frac{1}{2}\chi_E |z_3 M(\nabla.M)+z_4 [(M.\nabla)M]|^2
\label{Zequation}
\end{equation}
The spatial profile of the ground state is determined by the momentum dependent $\chi_M(q)$, the q for which it is a maximum. The ferromagnetic ground state would be unstable if the effect of the Lifshitz invariant were to move the ground state to finite q. Note however that the effective free energy for M, resulting from the Lifshitz invariant is quartic in M. In other words, a ferromagnet is stable as long as $b_M > (\chi_E(q)z)^2$. Since $\chi_E$ is divergent at T$_E$, a ferromagnetic ground state would be unstable near that temperature.

The thermodynamic fluctuations, which refer to the contributions of inhomogeneous terms as well as higher order interaction terms, lead to a qualitative change in all temperature dependent properties. There are two effects: the first is a downward shift in the critical temperature. The second is the movement of asymptotic mean field behavior, resulting in a different exponent. The temperature range of these effects can be encapsulated into the Ginzburg parameters.

The fluctuation effects arise from gradient terms, $\gamma_e (\nabla \P)^2$ and $\gamma_m (\nabla \M)^2$, in the free energy (Eq. \ref{freeenergy}). Near T$_{E0}$ the corresponding correlation length diverges. The subleading length scale derived from $\gamma_m$ starts to play an important role in the temperature range $T_M< T < T_{E0}$, but the relative role of $\gamma_m$ and $\gamma_e$ changes as one moves between T$_{E0}$ and T$_{M}$. Near T$_M$ there is an effective free energy similar to Eq. \ref{Zequation}, which includes $z_3$ and $z_4$, however a Gaussian calculation notes the effect of these terms as quartic in M and therefore negligible. These terms are also irrelevant in the RG sense in three dimensions, ($x\mapsto\frac{x'}{b}$), z$_i$ scales as $b^{2-d}$ in d-dimensions.

\begin{figure}[ht]
\includegraphics[width=90mm]{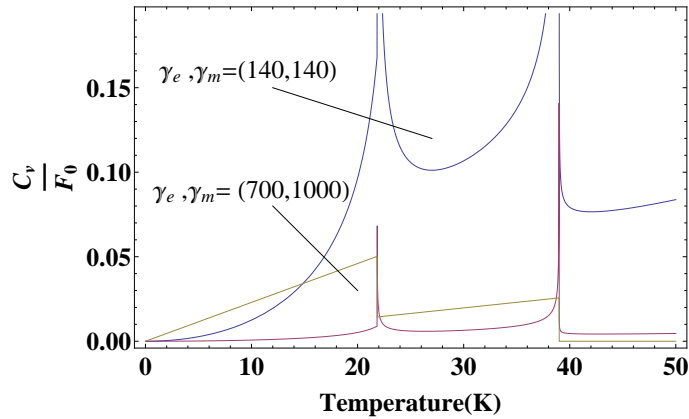}
\caption{ The most singular contribution to the specific heat from gaussian fluctuations, for different sets of correlation lengths}
\label{fluctuations}
\end{figure}

The fluctuation contribution to the specific heat was computed in the gaussian approximation, and shown in Fig. \ref{fluctuations}.

We first include gradients, $\gamma_e (\nabla \P)^2$ and $\gamma_m (\nabla \M)^2$ in the free energy. Then, we expand about the saddle point, $m_0 +\delta m$, $p_0 +\delta p$, and neglect cross-terms since this is meant to be an estimate. There are two independent contributions in that approximation, one for the gaussian integral over the polarization and one from the magnetization. The most singular terms in the specific heat take the approximate form
\[
C_{v\;fluct} \approx T^2 G(T) F(T)^{\frac{d}{2}-2}
\]
Where the functions G and F follow a pattern. Let $O_i$ denote the ith order parameter, $\M$ or $\P$. Then $G(T)=a/T_c+ 12 O_1 \frac{dO_1}{dT} + 2 k O_2 \frac{dO_2}{dT}$ and $F(T)=(\frac{T}{T_c}-1)+6b_1 O_1^2(T) + k O_2^2(T)$. The correlation lengths will take the form, $\xi_i^2=\frac{\gamma_i}{F(T)}$. The width of the critical region around either T$_c$ in three dimensions\cite{Goldenfeld} is given by the expression $T_G^{1/2}=\frac{k_B}{4 \Delta C_{mf} \xi_i^3}$, where $\Delta C_{mf}$ is the size of the jump in the specific heat from mean field theory.  We plot the behavior of these decoupled fluctuations in Fig. \ref{fluctuations}. It will take a measurement to determine to what extent the system can be treated within mean field theory. We expect the dispersion of polar phonons to be relatively weak, and relying on experience in ferroelectrics, the polar transition could even be first order.

\begin{figure}[htbp]
  \begin{center}
    \centering
    \includegraphics[width=3in, scale=0.5]{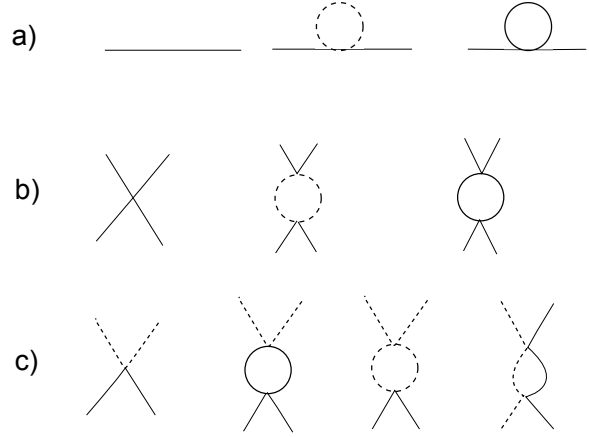} \quad
    \caption[]{The diagrams at one loop. a) two-point function b) vertex for one order parameter c) vertex coupling two order parameters. Dotted or solid lines refer to the propagators for each order parameter separately.}
    \label{diagrams}
  \end{center}
\end{figure}

It is instructive to consider the renormalization group flow and analyze the fixed points of our free energy. The critical behavior of two coupled order parameters each with $O(n)$ symmetry has already been computed in other contexts\cite{Kosterlitz}, and the results can be profitably reapplied in this new context. The necessary diagrams are shown in fig. \ref{diagrams}. The five flow equations, resulting from integrating out a shell of momentum $\frac{\Lambda}{b},\Lambda$,  are two copies of:
\begin{equation}
'r_i=b^2[r_i+ 4(n_i+2)u_i \int_{\frac{\Lambda}{b}}^{\Lambda} \frac{d^d q}{r_i+q^2}
+2 n_j k \int_{\frac{\Lambda}{b}}^{\Lambda} \frac{d^d q}{r_j+q^2} ]
\end{equation}
\begin{equation}
'u_i=b^{\epsilon}[u_i-4(n_i+8)u_i^2 \int_{\frac{\Lambda}{b}}^{\Lambda} \frac{d^d q}{(r_i+q^2)^2}
 - 4 n_j k^2 \int_{\frac{\Lambda}{b}}^{\Lambda} \frac{d^d q}{(r_j + q^2)^2}]
\end{equation}
and the flow for the coupling of the two order parameters:
\begin{equation}
'k=b^{\epsilon}[k-16 k^2 \int_{\frac{\Lambda}{b}}^{\Lambda} \frac{d^d q}{r_m+q^2} \frac{1}{r_e+q^2}
- 4(n_m+2)k u_m \int_{\frac{\Lambda}{b}}^{\Lambda} \frac{d^d q}{(r_m+q^2)^2}  -4(n_e+2)k u_e \int_{\frac{\Lambda}{b}}^{\Lambda} \frac{d^d q}{(r_e+q^2)^2} ]
\end{equation}

There are six fixed points determined by the flow of the quartic terms $u_i$ and k. In four of these, the order parameters decouple, $k=0$. The remaining two non-trivial ones, the Heisenberg-Heisenberg and so-called biconical fixed points, interchange their stability as a function of how many components the order parameters have. Fluctuation driven first order transitions\cite{HalperinLubenskyMaPRL} can occur if there is a runaway flow in the RG (when there is no fixed point accessible under RG iterations), and are absent in this model. At the double-Heisenberg fixed point, $u_{mc}=u_{ec}=k_c=\frac{\epsilon \Lambda^\epsilon 8 \pi^2}{n_e+n_m+8}$. Linearizing, the eigenvalues are all negative for $n_e+n_m \leq 4$ and $\epsilon>0$, but changes its stability if the components of the order parameter change. For the case n=3, that is for three dimensional vectors $\P$ and $\M$ considered here, the biconical fixed point is stable, and the double Heisenberg fixed point is unstable. If we have two easy planes instead and n=2, or very strong crystal field anisotropic such that n=1 effectively, then the double-Heisenberg fixed point is again stable. The anomalous dimension as in all one-loop scalar field theories vanishes, but we can anticipate corrections at higher order.

\section{Summary and Discussions}

We have studied the thermodynamic properties of a system with two order parameters and two transitions. Since, as a specific example, we have in mind a ferromagnet and a ferroelectric, the homogeneous interaction term is simply $M^2 P^2$. All the lowest order symmetry allowed interactions (without reference to crystallographic groups) were discussed in section IV. We calculate the phase diagram, order parameter temperature and field dependence, specific heat and susceptibility for the coupled order parameters. We follow through with a brief discussion of a renormalization group analysis.

Our results are (1) the cross susceptibility exists only below T$_M$, with a characteristic temperature dependence that diverges at T$_M$ in the thermodynamic limit. (2) The electric susceptibility changes discontinuously at T$_M$. The temperature T$_M(B,E)$ as a function of E follows a second order phase boundary. The effect of fluctuations (measured by the coefficients $\gamma_m$ and $\gamma_e$) would be to move T$_{E0}$ and T$_M$ lower by amounts depending on $\gamma_m$ and $\gamma_e$. For large $\gamma_m$ and $\gamma_e$ the fluctuations play insignificant roles.The fluctuation contributions in the vicinity of T$_{E0}$ are expected to be large. We expect though that qualitative features of this paper remain intact. The question remains whether the cross susceptibility is generated by fluctuations at least in the temperature range $T_M< T < T_{E0}$. There is no homogeneous factor that we can imagine which might lead to a nonzero $\chi_{12}$. The possibility of a defect generated $\chi_{12}$ remains a topic of future study.

\section{Acknowledgements}
We gratefully acknowledge discussions with A.Bhalla, A.B.Saxena, S.Obukhov, and R.Valdes-Aguilar. GRB was partially supported by by DOE DE-FG02-05ER46236. The work of SRP was supported by the National Science Foundation under Grant Number DMR-0426870.

\end{document}